\documentclass[reprint,superscriptaddress, amsmath,amssymb,pra, aps,floatfix,longbibliography]{revtex4-1}
\usepackage[utf8]{inputenc}
\usepackage[english]{babel}
\usepackage{hyperref}
\usepackage{dcolumn}
\usepackage{bm}
\usepackage{hyperref}
\usepackage[dvipsnames]{xcolor}
\usepackage{graphicx}
\usepackage[capitalise]{cleveref}
\usepackage{textcomp}
\usepackage{pdfcomment}
\usepackage{bold-extra}
\usepackage[normalem]{ulem}

\definecolor{gregRed}{RGB}{209, 0, 11}

\newcommand{\gregsout}{\bgroup\markoverwith{\textcolor{gregRed}{\rule[0.5ex]{2pt}{0.4pt}}}\ULon}

\DeclareUnicodeCharacter{2009}{\,}
\hypersetup{
 pdftitle={},    
 pdfauthor={Gregory Moille},     
 pdfsubject={DWsynthesis},   
 pdfcreator={Gregory Moille},   
 pdfproducer={LaTeX}, 
 pdfkeywords={Version 1.0},
 colorlinks=true,       
 linkcolor=blue,          
 citecolor=blue,        
 filecolor=blue,      
 urlcolor=blue}

\newcommand{\beginsupplement}{%
 \setcounter{table}{0}
 \renewcommand{\thetable}{S\arabic{table}}%
 \setcounter{figure}{0}
 \renewcommand{\thefigure}{S\arabic{figure}}%
 \renewcommand{\thesubsection}{S-\arabic{subsection}}
}
\usepackage{titlesec}
\titleformat*{\section}{\bfseries\Large}
\titleformat*{\subsection}{\bfseries}
\usepackage{caption}
\begin{document}


\title{Ultra-Broadband Kerr Microcomb Through Soliton Spectral Translation}

\author{Gregory Moille}
\email{gmoille@umd.edu}
\affiliation{Joint Quantum Institute, NIST/University of Maryland, College Park, MD, USA}
\affiliation{Microsystems and Nanotechnology Division, National Institute of Standards and Technology, Gaithersburg, MD, USA}
\author{Edgar F. Perez}
\affiliation{Joint Quantum Institute, NIST/University of Maryland, College Park, MD, USA}
\affiliation{Microsystems and Nanotechnology Division, National Institute of Standards and Technology, Gaithersburg, MD, USA}
\author{Jordan R. Stone}
\affiliation{Joint Quantum Institute, NIST/University of Maryland, College Park, MD, USA}
\affiliation{Microsystems and Nanotechnology Division, National Institute of Standards and Technology, Gaithersburg, MD, USA}
\author{Ashutosh Rao}
\affiliation{Microsystems and Nanotechnology Division, National Institute of Standards and Technology, Gaithersburg, MD, USA}
\affiliation{Institute for Research in Electronics and Applied Physics, University of Maryland,College Park, MD, USA}
\author{Xiyuan Lu}
\affiliation{Microsystems and Nanotechnology Division, National Institute of Standards and Technology, Gaithersburg, MD, USA}
\affiliation{Institute for Research in Electronics and Applied Physics, University of Maryland,College Park, MD, USA}
\author{Tahmid Sami Rahman}
\affiliation{Joint Quantum Institute, NIST/University of Maryland, College Park, MD, USA}
\author{Yanne Chembo}
\affiliation{Institute for Research in Electronics and Applied Physics, University of Maryland,College Park, MD, USA}
\author{Kartik Srinivasan}
\email{kartik.srinivasan@nist.gov}
\affiliation{Joint Quantum Institute, NIST/University of Maryland, College Park, MD, USA}
\affiliation{Microsystems and Nanotechnology Division, National Institute of Standards and Technology, Gaithersburg, MD, USA}

\date{\today}

\begin{abstract}
\noindent Broad bandwidth and stable microresonator frequency combs are critical for accurate and precise optical frequency measurements in a compact and deployable format. Typically, broad bandwidths (e.g., octave spans) are achieved by tailoring the microresonator's geometric dispersion. However, geometric dispersion engineering alone may be insufficient for sustaining bandwidths well beyond an octave. Here, through spectral translation induced by the nonlinear mixing between the soliton and a secondary pump, we greatly expand the bandwidth of the Kerr soliton microcomb far beyond the anomalous geometric dispersion region on both sides of the spectrum. We show that such nonlinear mixing can be summarized through the concept of synthetic dispersion, highlighting the frequency matching of the nonlinear process. Through detailed numerical simulations, we show that the synthetic dispersion model captures the system's key physical behavior, in which the second pump enables the non-degenerate four-wave mixing process of Bragg scattering, which spectrally translates the soliton and produces new dispersive waves on both sides of the spectrum, all while preserving low-noise properties across the full comb bandwidth. We experimentally demonstrate these concepts by pumping a silicon nitride microring resonator at 1063~nm and 1557~nm to enable the spectral translation of a single soliton microcomb so that a total comb bandwidth close to 1.6 octaves is realized (from 137~THz to 407~THz). We further examine the low-noise characteristics of the resulting comb state, through beat note measurements across the spectrum, measurements of the comb tooth spacing in both the primary and spectrally-translated portions, and use of an electro-optic comb to probe the region in which the primary and spectrally-translated comb portions overlap. Such ultra-broadband microcombs provide new opportunities for full microcomb stabilization in optical frequency synthesis and optical atomic clocks, while the synthetic dispersion concept can extend microcomb operation to wavelengths that are hard to reach solely through geometric dispersion engineering.
\end{abstract}

\maketitle

Microresonator frequency combs are promising for chip-scale metrology applications including coherent range measurements~\cite{riemensberger_massively_2020}, spectroscopy~\cite{picque_frequency_2019}, and optical clocks~\cite{yu_tuning_2019,newman_architecture_2019}. These applications are typically realized in the dissipative Kerr soliton (DKS) regime of microcomb operation~\cite{herr_dissipative_2016}, and often rely on stabilization of the comb repetition rate and carrier-envelope offset frequency, the latter usually through a $f$-2$f$ interferometer~\cite{udem_optical_2002,diddams_optical_2020}. $f$-2$f$ stabilization requires at least an octave of comb bandwidth, which can be achieved through geometric dispersion engineering~\cite{coen_modeling_2012, okawachi_bandwidth_2014} to create coherent dispersive waves (DWs) that broaden the comb spectrum~\cite{brasch_photonic_2016,li_stably_2017, pfeiffer_octave-spanning_2017,karpov_photonic_2018,yu_tuning_2019}. Although DWs significantly increase microcomb bandwidth, the power of these enhanced comb teeth is still orders of magnitude lower than the pump, so that the $f$-2$f$ technique remains challenging. For example, an end-to-end comb bandwidth of one octave is insufficient for self-referencing using the (centrally located) high-power pump. More complicated resonator cross-sections~\cite{moille_phased-locked_2018} and stacks of different materials~\cite{zhang_silicon_2012,dorche_advanced_2020} have been proposed to alter the dispersion in support of ultra-broadband combs, yet remain to be demonstrated experimentally. Other approaches for super-octave microcomb generation include combining $\chi^{(2)}$ and $\chi^{(3)}$ effects~\cite{chen_chaos-assisted_2020}, but such broadband combs usually present spectral gaps~\cite{szabados_frequency_2020}, and the suitability of such combs~\cite{chen_chaos-assisted_2020, hendry_experimental_2020, bruch_pockels_2020} for metrology has not been shown.\\
{\indent}Here, we present a low-noise microcomb whose span extends across 1.6 octaves - without spectral gaps - while bridging the telecom with near-visible wavelengths. This is made possible through dual pumping, in which the  second pump enables the $\chi^{(3)}$ process of four-wave mixing Bragg scattering (FWM-BS)~\cite{yulin_four-wave_2004,xu_cascaded_2013,li_efficient_2016} to significantly broaden the typical DKS state, by spectral translation of the soliton into other spectral bands. Using the dual pump scheme, we demonstrate that the DKS teeth, acting as the signal in the FWM-BS process, can be translated to new frequencies and effectively create new DWs on both sides of the original DKS spectrum, broadening its bandwidth by more than a factor two. The parametric nature of the FWM-BS process is such that phase coherence is expected to be maintained, which we probe experimentally through a series of noise measurements. In particular, heterodyne beat notes across the spectrum, measurements of the comb tooth spacing in both the original DKS portion and the spectrally translated portion, and a measurement of the relative noise between the comb teeth in the overlap between the two microcomb portions, are all consistent with the picture that FWM-BS spectrally translates the soliton -- thereby preserving its repetition rate -- to the spectral region surrounding the second pump, and the resulting 1.6 octave comb operates in a low-noise state. The incorporation of the FWM-BS spectral translation mechanism allows for a tunability and engineering of new DWs well beyond the limits imposed by geometric and material dispersion on conventional singly-pumped microresonator DKS states. To better understand the potential of this system, we introduce the new concept of synthetic dispersion, which captures the underlying physics and predicts the comb behavior as a function of resonator geometry and pump frequencies. Simultaneously, we perform a detailed numerical study using a single multi-pump Lugiato-Lefever Equation~\cite{chembo_spatiotemporal_2013,taheri_optical_2017} that accounts for the full set of $\chi^{(3)}$ processes occurring in the resonator and validates the novel concept of synthetic dispersion. The synthetic dispersion framework is further validated by close correspondence with our experimental measurements of ultra-broadband microcombs created by FWM-BS spectral translation of a DKS state.

%
%
\begin{figure*}[t!]
 \begin{center}
  \includegraphics[width=\textwidth]{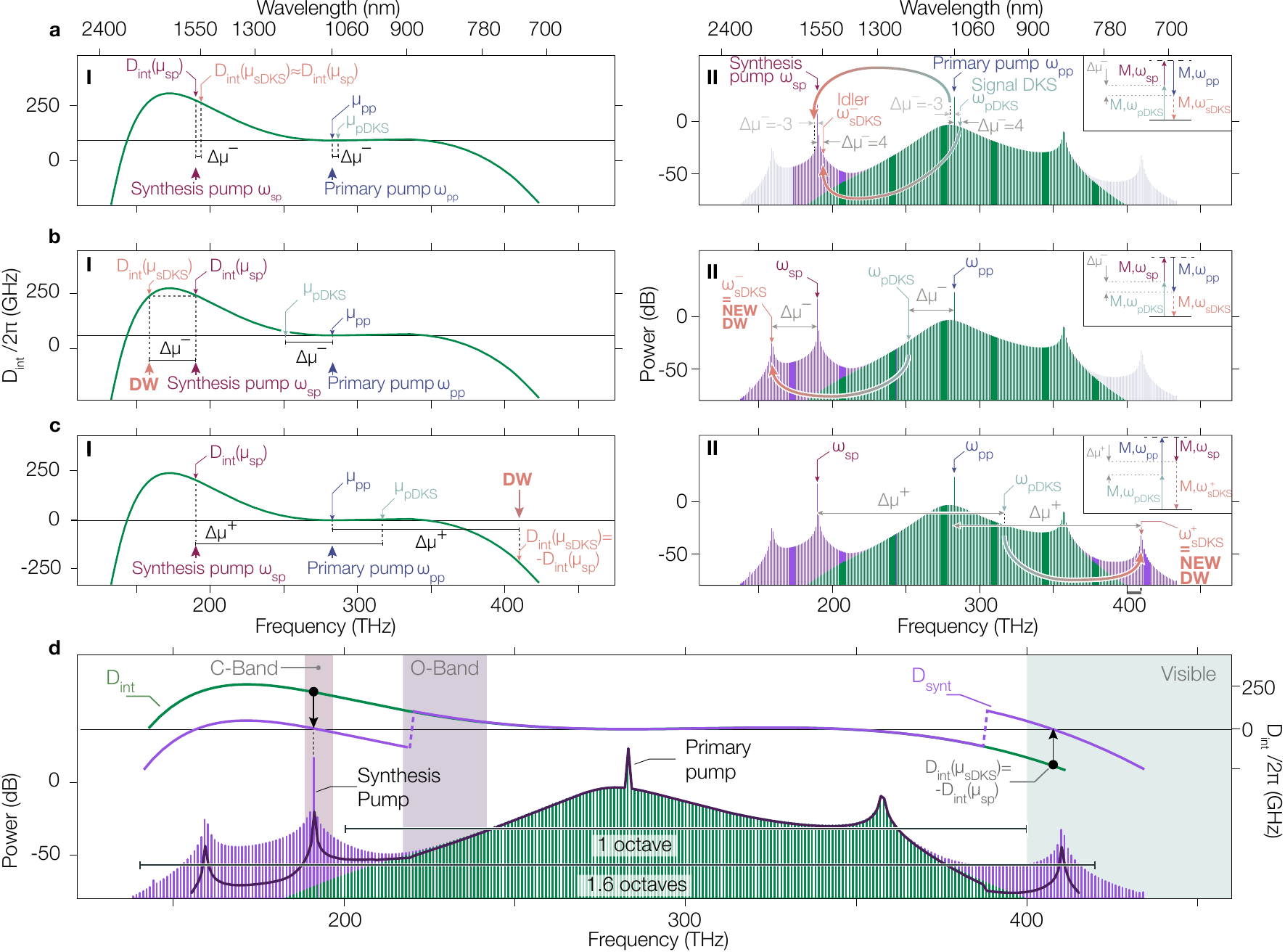}
 \end{center}
\vspace{0.1cm}
 \caption{\label{fig:ConceptSynthesis}\textbf{$|$~Principle of synthetic dispersion through four-wave mixing Bragg scattering (FWM-BS).} \textbf{a,} Integrated dispersion (panel I) of a ring resonator with two zero-crossings. Under driving by a single pump called the primary pump (pp), only one DW at high frequency is created (the potential low frequency DW is not generated due to power considerations), as shown in the green spectrum in panel II. This part of the spectrum is referred to as the primary portion. Inclusion of another pump, the synthesis pump (sp), allows for the FWM-BS condition to be respected, especially considering the negative (lower frequency) idler process (panel I). It effectively translates comb teeth surrounding the primary pump to spectral positions surrounding the synthesis pump, creating a synthesized portion of the frequency comb (panel II, purple spectrum). \textbf{b,} The same negative idler frequency and momentum conservation condition is met for larger detuning between the primary pump and the signal, as the higher order dispersion coefficients allow for a roll-off of the integrated dispersion, and therefore another mode at the same integrated dispersion value as at the synthesis pump exists (panel I). Such efficient FWM-BS effectively creates a new DW at low frequency (panel II). \textbf{c, } Now considering the higher frequency idler due to FWM-BS, both the the mode number matching condition and frequency matching conditions are changed (panel I). The frequency matching condition now imposes that the integrated dispersion at the idler must be of equal and opposite sign to the value at the synthesis pump, and therefore a DW in the primary portion of the comb must already exist (e.g., at 355~THz) to allow a change of sign of $D_\mathrm{int}$ (panel I). In our case, this frequency matching condition is met twice, but the mode matching condition permits only the high-frequency mode to undergo FWM-BS. In the same manner as previously, it effectively creates a new DW at high frequency (at 410~THz), extending the frequency comb bandwidth on this side of the spectrum (panel II). \textbf{d,} Introduction of a synthetic dispersion $D_\mathrm{synth}$ (purple, right axis) that captures the nature of the translated portions as new DWs at low and high frequencies and exhibits a clear difference from the integrated dispersion of the resonator (solid green, right axis). The resulting LLE simulated comb spectrum with a single pump (green) and a dual pump (purple) exhibit a clear difference of bandwidth. The synthetic integrated dispersion emulates the dual pump system by an effective single pump system (dark purple line) and shows close agreement with the dual pump simulation. 0~dB is referenced to 1~mW, i.e., dBm.
}
\end{figure*}%

\noindent \textbf{Spectral translation and the synthetic dispersion framework}--A microresonator's integrated dispersion $D_\mathrm{int}$ represents the variation of the cold cavity resonance frequencies away from an equidistant frequency grid (\textit{i.e.} the DKS comb teeth) spaced by $D_1/2 \pi$, the free spectral range (FSR) around the primary pump~\cite{kippenberg_dissipative_2018},  hereafter annotated with the label $pp$. $D_\mathrm{int}(\mu) = \omega_\mathrm{res}(\mu) - \omega_\mathrm{\textsc{dks}}(\mu)  = \omega_\mathrm{res}(\mu) -  \left(\omega_\mathrm{pp} + D_1 \mu \right)$, with $\mu$ defined as the mode order relative to the pumped mode $\mu_\mathrm{pp} = 0$ (\textit{i.e.} $D_\mathrm{int}(\mu_\mathrm{pp})=0$), $\omega_\mathrm{res}(\mu)$ being the cavity resonance frequency of the mode $\mu$, and $\omega_\mathrm{DKS}(\mu)$ being the $\mu^{\text{th}}$ DKS comb tooth frequency. Hence, when the cavity resonances match the DKS comb teeth frequencies, \textit{i.e.}, $D_\mathrm{int}(\mu_\mathrm{\textsc{dw}})=0$, a resonant enhancement happens, leading to the DW creation at the mode position $\mu_\mathrm{\textsc{dw}}$ (\cref{fig:ConceptSynthesis}a).
In this case (and for the microresonators we study below),  $D_\mathrm{int}$ is such that only one primary DW ($DW'$) is created, as the second zero crossing (on the low frequency side) is too far from the primary pump to yield appreciable energy. \\
{\indent}The use of an auxiliary pump enables straightforward access to soliton states, through an effective temperature compensation mechanism that bypasses the thermal bistability~\cite{lu_deterministic_2019, zhou_soliton_2019,zhang_sub-milliwatt-level_2019}. Moreover, simultaneous spectral broadening of the comb, attributed to cross-phase modulation (XPM) effects, has also recently been observed~\cite{zhang_spectral_2020}, though the magnitude of the effect was limited. Here, we consider a dual-pumped system in a regime where much more significant spectral broadening is realized. We pinpoint the origin of the strong increase in comb bandwidth as originating from an interband four-wave mixing Bragg scattering (FWM-BS) process. FWM-BS is mediated by the combination of a strong, secondary pump (hereafter referred to as the synthesis pump $sp$) and the primary pump $pp$, and results in phase-coherent spectral translation of comb teeth across wide spectral gaps determined by the difference in pump frequencies. While FWM-BS of a single frequency continuous wave input has been demonstrated in a microcavity~\cite{li_efficient_2016}, and its role in soliton-DW mixing in the context of optical fibers has previously been studied in intraband~\cite{xu_cascaded_2013} and interband cases~\cite{yulin_four-wave_2004}, here we show how it can play a critical role in the creation of ultra-broadband microresonator frequency combs.

We consider a FWM-BS framework where the signal can be any comb tooth of the primary soliton ($pDKS$) resulting from the primary pump $pp$ (i.e., the comb that would be obtained through a single pump, called the primary portion), and is converted into an idler that is another spectral component of the comb - hereafter called the synthesized portion ($sDKS$) - through application of the synthesis pump $sp$. This process must respect the fundamental criteria of energy and momentum conservation, which in a ring resonator translate to frequency matching and azimuthal mode number matching $\omega_\textsc{sDKS}  = \omega_\mathrm{pDKS} \pm |\omega_\textsc{pp} - \omega_\mathrm{sp}|$ and $\mu_\textsc{sDKS}  = \mu_\mathrm{pDKS} \pm |\mu_\textsc{pp} - \mu_\mathrm{sp}|$, respectively. Using the integrated dispersion previously defined, these fundamental conditions can be summarized in a single equation (see Extended Data~\cref{sub:sup-mat-FWMBS}.)

\begin{align}
    \label{eq:dint-fwm-general}
    \Big (\mu_\mathrm{pDKS} \pm \mu_\mathrm{pp} \Big)D_1 =
    & \Big( D_\mathrm{int}(\mu_\mathrm{sDKS}^{\pm}) \pm D_\mathrm{int}(\mu_\mathrm{sp}) \Big) + \nonumber \\ &  \Big( \mu_\mathrm{sp} \pm \mu_\mathrm{sDKS}^{\pm} \Big)D_1
\end{align}
This results in a simple pair of conditions for FWM-BS based on the idler that is considered:
\begin{align}
    &\begin{cases}
        \label{eq:FWM-BS-pair-m}
        \Delta D_\mathrm{int}^{-} =  D_\mathrm{int}(\mu_\mathrm{sDKS}^{-}) -  D_\mathrm{int}(\mu_\mathrm{sp}) = 0\\
        \Delta \mu^{-} =  \mu_\mathrm{sDKS}^{-} - \mu_\mathrm{sp} = \mu_\mathrm{pDKS} - \mu_\mathrm{pp}
    \end{cases}\\
    &\begin{cases}
        \label{eq:FWM-BS-pair-p}
        \Delta D_\mathrm{int}^{+}  = D_\mathrm{int}(\mu_\mathrm{sDKS}^{+})  + D_\mathrm{int}(\mu_\mathrm{sp}) = 0\\
        \Delta \mu^{+} = \mu_\mathrm{sDKS}^{+} - \mu_\mathrm{pp} = \mu_\mathrm{pDKS} - \mu_\mathrm{sp} \\
    \end{cases}
\end{align}
Here the superscript $\pm$ denotes the two kinds of idler that can result from a FWM-BS process, one at higher frequency (+) and the other at lower frequency (-) than the signal. Due to phase matching considerations,  only one of each would match the FWM-BS condition for a given signal.

The first case to consider is for a signal that is close to the main pump (\cref{fig:ConceptSynthesis}a). Only the low frequency idler satisfies the condition in \cref{eq:FWM-BS-pair-m}. Therefore, the FWM-BS process translates a DKS comb tooth close to the primary pump into a comb tooth close to the synthesis pump. As the mismatch in the integrated dispersion $\Delta D_\mathrm{int}^{-}$ (i.e. mismatch in the fundamental energy conservation of the FWM process) increases with the mode spacing $\Delta \mu^{-}$, the efficiency of the FWM-BS process decreases, giving rise to the Lorentzian spectral shape around the synthesis pump. However, comb teeth close to the primary pump are not the only ones that can be efficiently spectrally translated. Due to the higher order dispersion coefficients leading to a roll-off of the integrated dispersion, the energy conservation condition $\Delta D_\mathrm{int}^{-} = 0$ is met again for large enough mode spacing (\cref{fig:ConceptSynthesis}b)). Therefore, the mode frequency exhibiting the same integrated dispersion value as at the synthesis pump would exhibit efficient FWM-BS. This process extends the frequency comb bandwidth on the low frequency side, with a new DW generated near 160~THz. The system we consider presents a large asymmetry in the integrated dispersion, and therefore the condition $\Delta D_\mathrm{int}^{-} = 0$ is only met once. However, even for symmetric integrated dispersion in which there is an additional $\Delta D_\mathrm{int}^{-} = 0$ on the high frequency side, no corresponding high frequency DW will be generated through this negative idler process. This is because the phase matching criterion from Eq.~\ref{eq:FWM-BS-pair-m} requires the mode spacing between the synthesis pump and the synthesized portion to match the separation between the primary pump and the primary portion, and such a separation is too large for the primary portion to contribute an adequate power signal for the FWM-BS process.

That being said, Eq.~\ref{eq:FWM-BS-pair-p} indicates that a higher frequency idler can be generated for any modes whose integrated dispersion is equal and opposite in sign to the integrated dispersion value at the synthesis pump. A comb tooth from the synthesis portion $\mu_\mathrm{sDKS}^{+}$ must respect momentum conservation, which in this case means that the mode spacing between the signal (comb tooth from the primary portion) and the synthesis pump must match the mode spacing between the idler (comb tooth from the synthesized portion) and the primary pump. Once again, due to the higher order dispersion that allows a zero crossing of the integrated dispersion, negative values of $D_\mathrm{int}$ are possible, and an efficient FWM-BS process can happen at frequencies beyond the original high-frequency DW of the single pumped DKS. This results in a new DW near 409~THz, extending the bandwidth of the frequency comb to higher frequencies, i.e. toward the visible. Interestingly, another frequency matching condition in our case would occur at low frequency (near 138~THz), as another zero crossing of the integrated dispersion happens on this side of the spectrum. However, momentum conservation is not respected, as the idler must have a higher mode number than the signal, and therefore no FWM-BS happens here.

\begin{figure*}[t!]
 \begin{center}
  \includegraphics[width=\textwidth]{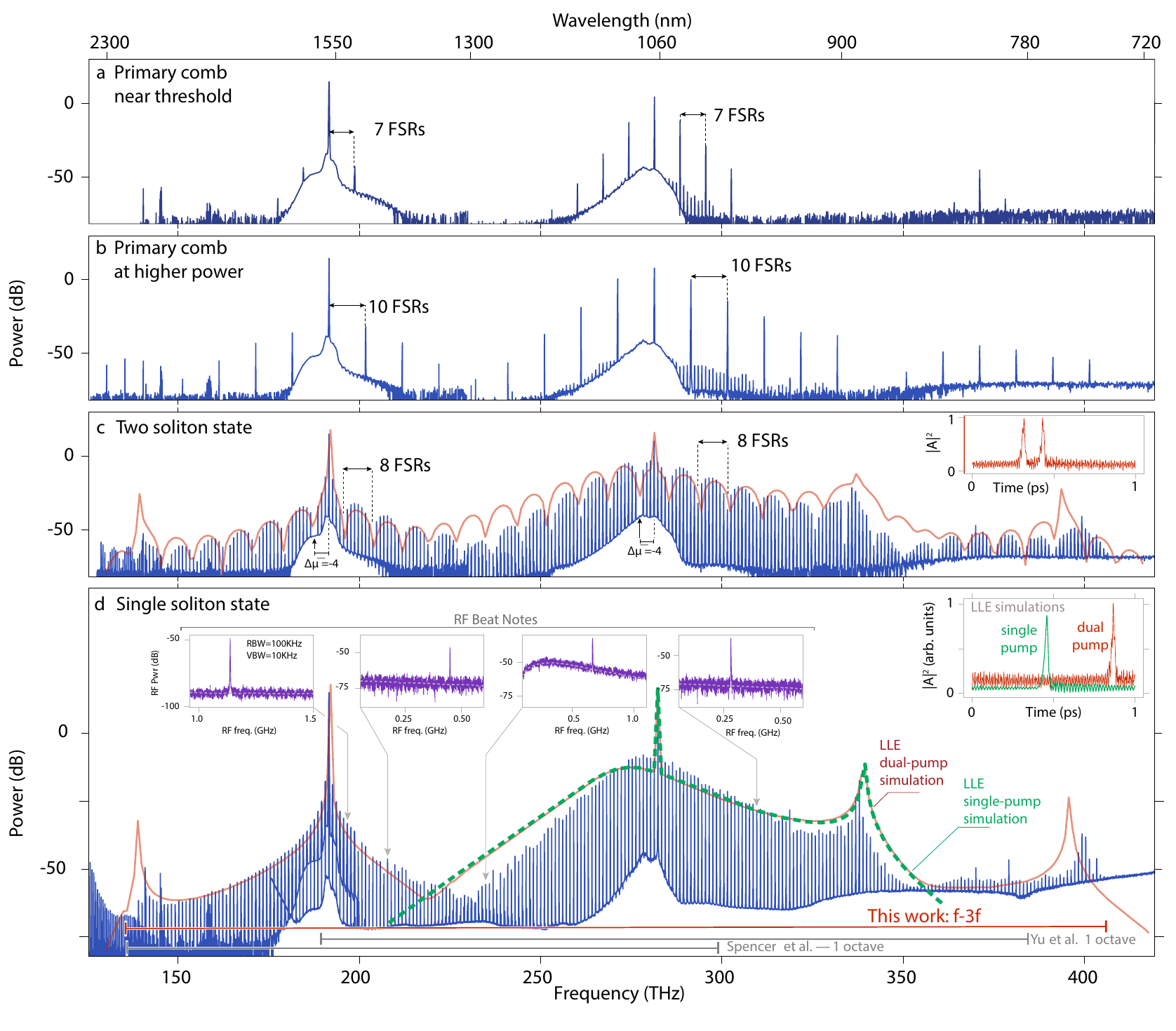}
 \end{center}
 \caption{\label{fig:ExperimentalTwoOCtave}\textbf{$|$~Spectral translation to create ultra-broadband microcombs}. A microring resonator with $RW$=1117~nm is pumped by a primary pump at 282~THz and synthesis pump at 192~THz. The comb behavior surrounding the primary pump is spectrally translated to the synthesis pump portion. \textbf{a,} Primary comb generation with low primary pump power near threshold. The comb spacing is equal to seven free spectral ranges (FSRs) and this spacing is reproduced around the synthesis pump, where idlers are created seven FSRs away from the synthesis pump, highlighting the mixing process between the two pumps and the primary portion comb teeth. \textbf{b, }Primary comb generation at a higher primary pump power where, as previously, the spectral spacing in the primary portion is matched by that in the synthesis portion, as expected by the FWM-BS theory. \textbf{c} Two-soliton state, where the characteristic 8 FSR modulation in the comb envelope is replicated near the synthesis pump. The inset shows the LLE-calculated two-soliton pulse arrangement that results in the simulated comb envelope shown in red. We highlight the missing comb tooth in the primary portion ($\Delta\mu$=-4), whose absence is translated onto the synthesized portion of the comb, respecting the FWM-BS phase matching condition. \textbf{d,} Single soliton state, where the impact of the synthesis pump is to expand the comb bandwidth to 1.6 octaves and create new DWs on both ends of the spectrum. The spectrum agrees with the generalized LLE solution using the dual-pump model (red line), and greatly exceeds the expected spectrum if just the primary pump is applied (dashed green line). The phase coherent nature of the generated comb is further verified through beat note measurements with narrow linewidth lasers throughout the comb spectrum (four left insets). The noise floor for each measurement is shown in dashed lines, and is higher in the O-band due to use of an additional RF amplifier. The rightmost inset shows the LLE simulation of the expected time-domain behavior under dual pumping (red) and if only the primary pump is applied (green). The horizontal bars at the bottom of the graph compare the span achieved here with octave-spanning DKSs from Refs.~\cite{spencer_optical-frequency_2018,yu_tuning_2019}. We note that the low frequency portion of the spectrum exhibits OSA artefacts, at 146~THz, 159~THz, and $<$135~THz; the shortest DW at 141~THz is not impacted by these artefacts. 0~dB is referenced to 1~mW, i.e., dBm.}
\end{figure*}%

Using this fundamental property that FWM-BS translates the primary DKS comb teeth into new spectral regions while maintaining the comb tooth spacing, we introduce $D_\mathrm{synt}$ (\cref{fig:ConceptSynthesis}d), a synthetic dispersion that captures the essence of the FWM processes we have presented. $D_\mathrm{synt}$ is an effective integrated dispersion that incorporates the combined influence of the geometric integrated dispersion and the synthesis-pump-induced FWM-BS processes. $D_\mathrm{synt}$ is essentially a piece-wise shifted version of $D_\text{int}$, with the two being equal in the spectral region surrounding the primary pump, and differing in the regions where FWM-BS causes a broadening of the spectrum and the generation of new DWs. In these regions, we simply use the FWM-BS conditions for creating the new DWs to determine how to shift $D_\text{int}$ (either up or down) so that its zero crossings are appropriately located. Stitching together the different regions of the synthetic dispersion is accomplished by taking into account FWM-BS power considerations, and in particular, where FWM-BS is less efficient. In our case, this is at the midpoint between the primary pump and the synthetic pump, and in between the primary DW (i.e., that generated by the original DKS) and the high frequency FWM-BS DW. This approach provides a linear approximation of where the DWs will be created ($D_\mathrm{synt}$=0) and helps estimate the spectral extent of the frequency comb. To test its validity, we simulate the behavior of the system using a generalized version of the Lugiato-Lefever Equation (LLE), described in detail in the Extended Material~\cref{sup-eq:LLE}. This version of the equation has not been subjected to typical simplifications, and in particular, the pumped modes remain as phase terms relative to the center of the frequency domain, so that the evolution of the intracavity electric field under multiple driving fields can be studied~\cite{taheri_optical_2017}.

%
%
%

\noindent \textbf{Ultra-broadband microcombs} -- To experimentally study the above phenomena, we perform measurements on Si$_3$N$_4$ microring resonators whose design and basic characterization are described in the Methods and Extended Data. The resonators are pumped in two bands, with a primary pump around 1063~nm and a synthesis pump around 1557~nm, and the coupling enabled by a tailored pulley waveguide geometry that realizes a relatively flat coupling rate across a wide spectral range~\cite{moille_broadband_2019} (see Extended Data~\cref{sub:sup-coupling}). We first show the spectral translation nature of the dual pump system for different microcomb states. Figure~\ref{fig:ExperimentalTwoOCtave}a-b shows the spectral behavior for primary combs generated just above threshold and at a higher power, respectively, where in both cases we observe that the comb tooth spacing (7~FSR and 10~FSR, respectively) surrounding the primary pump is retained around the synthesis pump and the higher frequency region between 350~THz to 400~THz. This behavior persists as we reach the soliton regime, in which the synthesis pump provides both a new nonlinear mixing mechanism (FWM-BS) as well as thermal stabilization, with clear signatures of soliton steps observed (extended data \cref{fig:setup}). Figure~\ref{fig:ExperimentalTwoOCtave}c shows the results for a two-soliton state, and in Fig.~\ref{fig:ExperimentalTwoOCtave}d, spectral translation of a single soliton state and the generation of additional DWs that greatly expand the comb spectrum is demonstrated. In each of these states, the spectral separation between the comb lines remains the same between the primary portion and the synthesized one, illustrating that the comb lines from the synthesized portion are unlikely to be due to the synthesis pump alone, and instead are a result of the mixing between both pumps and the primary portion comb lines. In addition, the clear translation of the two-soliton comb envelope modulation pattern onto the synthesis component confirms this point. This spectral pattern (due to the relative phase of the two pulses circulating in the cavity) effectively acts as a modulation of the signal power in the FWM-BS process, and the replication of this pattern in the synthesized portion of the comb follows the expectation for FWM-BS that the generated idler power is linearly proportional to the input signal power. For example, a comb tooth 4~FSRs below the primary pump ($\Delta\mu=-4$) is absent (likely due to an avoided mode crossing), and this is replicated by the absence of a comb tooth 4~FSRs below the synthesis pump. This also emphasizes the phase-matching condition and is consistent with the FWM-BS framework described previously.

\begin{figure}[t!]
 \begin{center}
  \includegraphics[width=\linewidth]{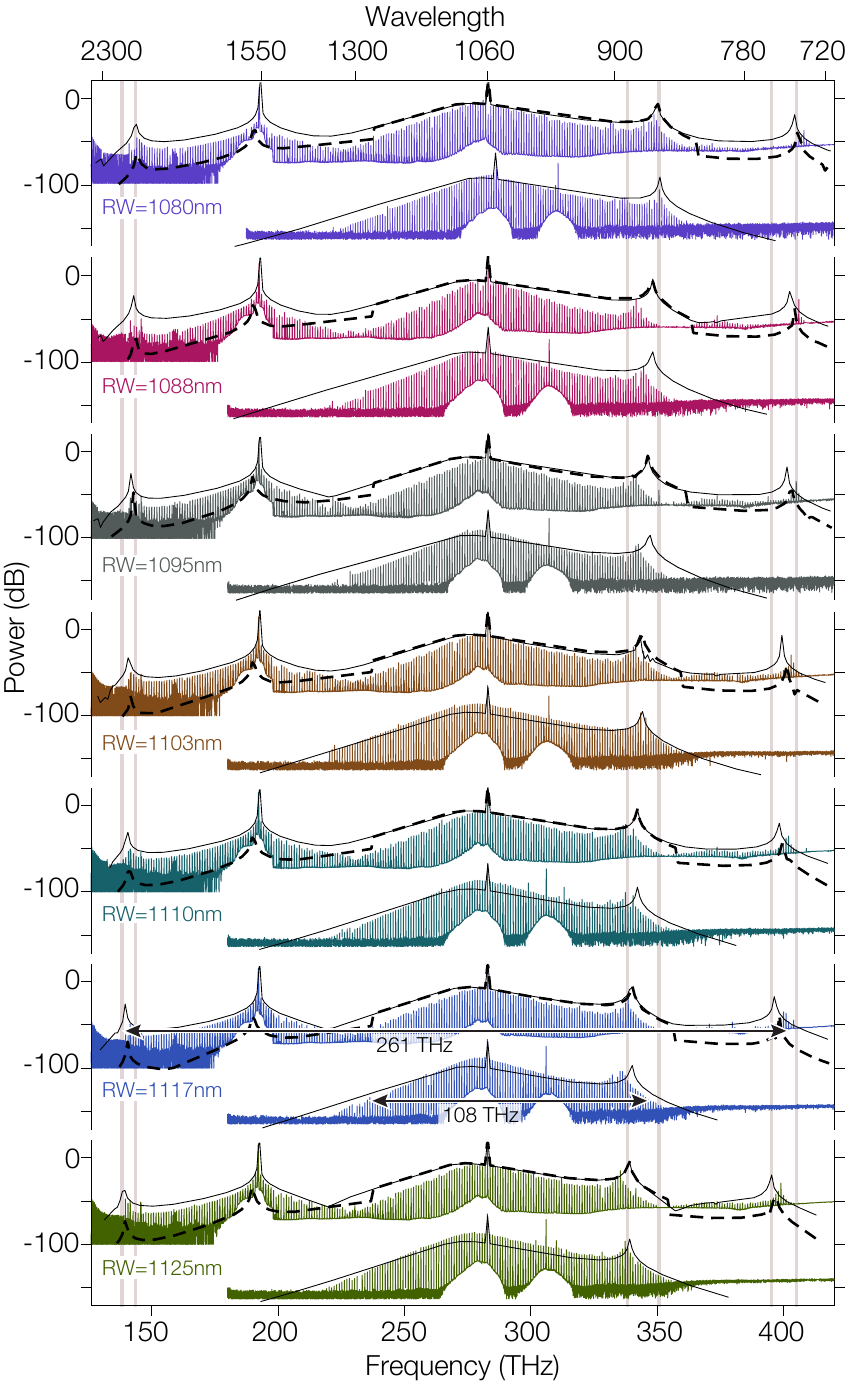}
 \end{center}
 \caption{\label{fig:CombGeomDependence}\textbf{$|$~Ultra-broadband soliton spectral translation: geometric dependence.} Geometric dispersion impacts the location of the generated DWs, much like the case in singly-pumped DKS devices. Here, microrings are pumped at 192~THz and 282~THz with a pump power of 200~mW and 250~mW respectively, for ring widths (RWs) from 1080~nm to 1125~nm. In each case, an ultra-broadband microcomb is generated in which the soliton comb teeth surrounding the primary pump are spectrally translated by the synthesis pump. Single pump DKS states are also showcased for each RW, where thermal stability was obtained through cross-polarized counterclockwise pumping at 305~THz. The solid black lines represent the expected single soliton spectra calculated through the dual-pumped generalized LLE and single-pump LLE for their corresponding counterpart experiments. The dashed lines represent the microcomb envelope prediction using the synthetic dispersion and single-pump LLE. The light grey solid line are visual guides showcasing the shift of the DWs with RW. 0~dB is referenced to 1~mW, i.e., dBm.}
\end{figure}%
\begin{figure*}[!t]
 \begin{center}
  \includegraphics[width=\linewidth]{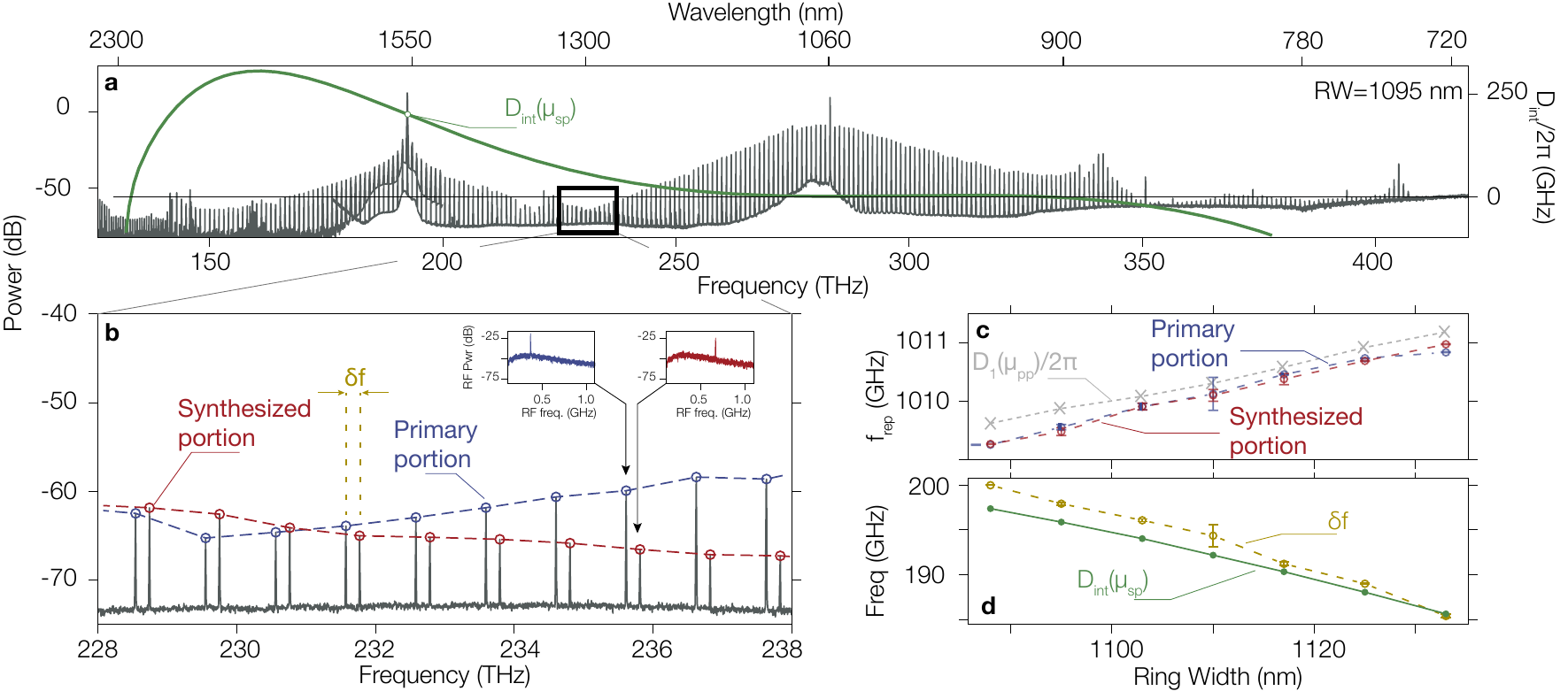}
 \end{center}
 \caption{\label{fig:OverlapMeasure}\textbf{$|$~Overlap of the primary and synthesized portions of the comb spectrum.} \textbf{a,} Spectrally-translated single DKS spectrum with the highlighted overlap region. The integrated dispersion computed at the primary pump frequency (282~THz) is shown in green, with an annotation highlighting the expected frequency offset between the DKS primary portion comb tooth and the cavity resonance at the synthesis pump frequency. \textbf{b,} Zoom-in of the overlap region, highlighting the primary portion of the DKS (blue) and the synthesized portion (red). The beat note of pairs of adjacent comb teeth separated by the overlap-offset frequency $\delta f$ are displayed in the insets. \textbf{c,} Measured comb tooth spacing for both the primary and synthesized portion in the overlap region, exhibiting a uniform value across the spectrum that is ring width dependent. The linear FSR around the main pump is measured and reported in the gray dashed line, and is close to the measured repetition rate of the DKS. The error bars represent the variance of the measured $f_\mathrm{rep}$ for different pairs of comb teeth. \textbf{d,} Measurement of the overlap-offset frequency $\delta f$ and the predicted value of the integrated dispersion about the primary pump, evaluated at the synthesis pump frequency ($D_\mathrm{int}(\mu_\mathrm{sp})$). The overlap-offset frequency is uniform across the overlap region, and is well-predicted by $D_\mathrm{int}(\mu_\mathrm{sp})$, indicative of the intrinsic detuning between the DKS teeth and the cavity resonances frequencies. The error bars represent the variance of the measured $\delta f$ for different pairs of comb teeth. 0~dB is referenced to 1~mW, i.e., dBm.}
\end{figure*}

When reaching the single soliton state, the comb extends from 137~THz to 407~THz, a span allowing f-3f stabilization  and a significant increase in bandwidth relative to state-of-the-art DKS microcombs~\cite{spencer_optical-frequency_2018,yu_tuning_2019}. The comb envelope is in good agreement with the predictions of the generalized LLE model, which incorporates both the primary and synthesis pumps, in terms of the overall comb envelope, the spectral positions of the different DWs, and the $>$80~dB dynamic range in comb tooth power across the ultra-broadband spectral range. Finally, we note that the LLE provides insight about the nature of the intracavity field (\cref{fig:ExperimentalTwoOCtave}c-d, insets). In the time domain, it predicts a two soliton pulse (Fig.~\ref{fig:ExperimentalTwoOCtave}c) and a single soliton pulse (Fig.~\ref{fig:ExperimentalTwoOCtave}d) with an estimated pulse duration (full-width at half-maximum) of 16~fs. The pulses sit on a modulated background characteristic of DWs on both sides of the spectrum, and the pulse itself shows some amount of structure. In contrast, we also plot the expected time-domain behavior for a singly-pumped soliton state with a single DW (Fig.~\ref{fig:CombGeomDependence}d), where the background modulation is larger on one side of the pulse, and the pulse itself shows no additional structure. This highlights the continuous wave nature of the translated portion of the spectrum, which does not create new pulses, but rather increases the background modulation in the same manner as DWs do. We note that the pulsed behavior from the time-domain simulations is consistent with coherence across the whole 1.6 octave bandwidth of the comb. \\
{\indent}To probe the comb coherence in the single soliton case, we perform beat note measurements with narrow linewidth tunable lasers (at 970~nm, 1270~nm, 1420~nm, and 1520~nm) positioned at different locations within the comb spectrum, covering many different parts of the comb with different spectral shapes. In each case (inset to \cref{fig:ExperimentalTwoOCtave}c), the beat note is a single tone, which is a signature of the phase-coherent nature of the frequency comb~\cite{herr_universal_2012, raja_electrically_2019} and is to be contrasted with the multiple beat notes that might be expected for modulation instability processes that lead to sub-comb formation. The beat notes at 1270~nm and 970~nm evaluate the primary portion of the DKS, and are thus expected to exhibit clear single tones, reflecting the single pulse nature of the DKS in the resonator. The beat notes at 1420~nm and at 1520~nm, which lie in spectrally-translated portion near the synthesis pump, support the hypothesis of the binding of the synthesis pump with the single soliton and the coherence of the synthesized component of the comb. Later, we will strengthen these conclusions through heterodyne measurements between the two portions of the comb.\\

%
%
%
\noindent \textbf{Soliton spectral translation: geometric dependence} — To explicitly demonstrate the impact of the synthesis laser on spectral translation of soliton microcombs, in Fig.~\ref{fig:CombGeomDependence} we compare spectra generated when both primary and synthesis pump lasers drive nonlinear processes in a series of microresonators to the case where only the primary pump drives soliton generation. Thermal stabilization in the latter case is achieved by a counterpropagating cross-polarized laser at 980~nm; the opposite propagation direction and orthogonal polarization ensure that its impact on nonlinear dynamics is minimized. For each microring, we see that effect of the synthesis laser is to mediate spectral translation and new DW generation, while leaving the portion surrounding the original soliton state essentially unchanged, as predicted from \cref{fig:ConceptSynthesis}. The microrings differ only in their ring widths, which range from 1080~nm to 1125~nm, and since the microring cross-section strongly influences the integrated geometric dispersion and the resulting synthetic dispersion under dual pumping, we expect this $RW$ variation to impact the generated comb spectra and the DW positions. Each device exhibits an ultra-broadband spectrum, and as expected, all DWs tune with ring width. The generalized LLE described in the Extended Data~\cref{sub:sup-lle} provides good agreement (solid black lines in Fig.~\ref{fig:CombGeomDependence}) with the obtained experimental spectra and reproduces the observed DW tuning. In addition, the single-pumped synthetic dispersion simulations (dashed black lines in \cref{fig:CombGeomDependence}) provide similar predictions for the comb envelope and DW positions, highlighting the utility of our heuristic model and its use as a predictive tool to design ultra-broadband frequency combs. However, it is important to note that both the single-pump LLE and the more generalized $N$-pump LLE rely on the basic assumption of a single fixed frequency grid, indexed by the mode number $\mu$, through Fourier transform of the fast time temporal profile. Hence these models do not capture any frequency discrepancy between the DKS teeth and the nearest cavity resonances. 


\begin{figure*}[!t]
 \begin{center}
  \includegraphics[width=\linewidth]{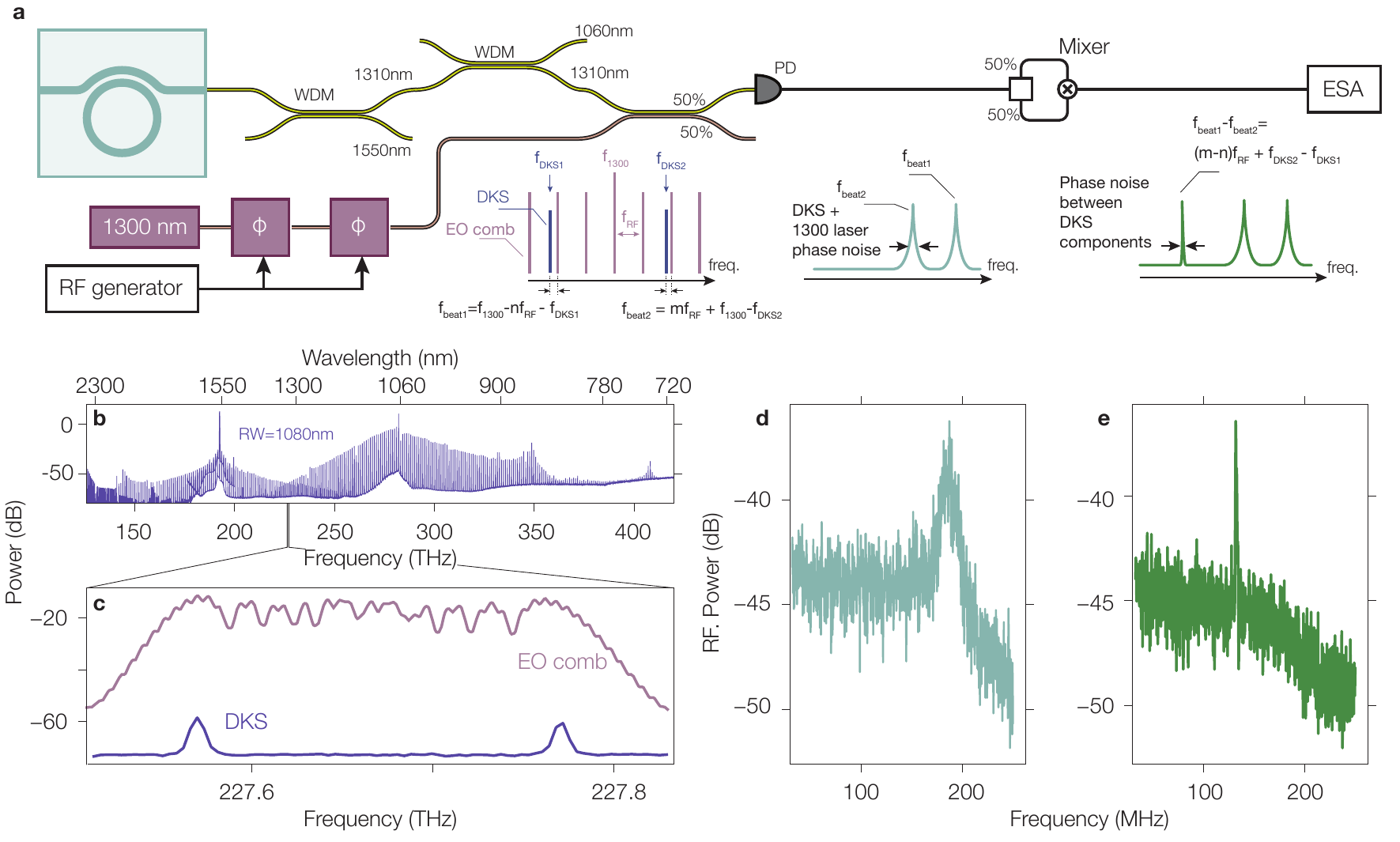}
 \end{center}
 \caption{\label{fig:BeatNoteStitching}\textbf{Relative phase noise between the primary and synthesis portions.} \textbf{a,} Schematic of the measurement setup through which the relative phase noise of two comb teeth in the stitching region, one from the primary comb portion and one from the synthesis comb portion, is measured with the assistance of a 1300~nm EO comb. \textbf{b,} Spectrum of the microcomb device under investigation. \textbf{c,} Two microcomb teeth from the stitching region (blue) plotted along with the EO comb (purple). The EO comb teeth are not fully resolved due to the limited spectral resolution of the optical spectrum analyzer in comparison to the EO comb repetition frequency (6.0156~GHz). \textbf{d,} Beat note of the EO comb with the two microcomb teeth. \textbf{e,} Mixing of the two beat notes produced between the EO comb and the microcomb teeth to retrieve the relative phase noise of the microcomb teeth, without the (free-running) 1300~nm laser phase noise contribution.}

\end{figure*}

\noindent \textbf{Composite frequency comb overlap frequency offset} — Closer inspection of ~\cref{fig:ExperimentalTwoOCtave}d and every spectrum in \cref{fig:CombGeomDependence} reveals an important feature: the primary portion and the synthesized portion do not overlap perfectly, resulting in a composite frequency comb. Although we are pumping the same mode family in both bands, the overlap region in the comb spectrum (\cref{fig:OverlapMeasure}a and b) exhibits pairs of adjacent comb teeth with an overlap-offset frequency $\delta f = f_\mathrm{sDKS} - f_\mathrm{pDKS}$ that is smaller than the repetition rate of the DKS. We now consider whether $\delta f$ remains the same across this `stitching’ region.\\
{\indent}We characterize the comb tooth spacing, \textit{i.e.} the repetition rate $f_\mathrm{rep}$ in the primary and synthesized portions of the comb, for the state shown in \cref{fig:OverlapMeasure}a. This is done by measuring beat notes between a helper laser and the comb teeth, with the helper laser wavelength determined by a wavemeter (accuracy $\approx$~50~MHz). Through measurement across the helper laser tuning range (228~THz to 238~THz), we find that (\cref{fig:OverlapMeasure}b) $f_\mathrm{rep}$ is uniform and equal for both the primary and synthesized portions, and its value is close to the FSR around the primary pump (equivalently, $\delta f$ is uniform throughout the overlap region). We then repeat measurement of $f_\mathrm{rep}$ and $\delta f$ for devices with different $RW$, and find that these conclusions hold, with the specific measured values dependent on $RW$ (Fig.~\ref{fig:OverlapMeasure}c-d). These conclusions are understood by the fact that the FWM-BS process must respect frequency matching, and thus the soliton repetition rate will be transferred to the synthesized portion of the comb.\\
{\indent}In addition, contrary to the strong coupling in a dual pump system that has been theoretically studied~\cite{taheri_optical_2017, hanson_bichropumping}, the synthesis pump here does not change the DKS repetition rate or shift the frequency of resonance enough to enter the regime where Arnold tongues and sychronization might exist~\cite{skryabin_arnoldtongues}. We believe that such locking cannot happen in our devices because the overlap frequency shift cannot be compensated by the Kerr shift at the synthesis pump ($\approx -1.3$~GHz). Thus, this overlap-offset frequency between the primary and synthesized portions of the comb can be understood as the fundamental discrepancy between the soliton comb tooth frequency and the cavity resonance frequency~\cite{guo_universal_2017}. As such, we expect that $\delta f$ should be given by the value of the primary pump integrated dispersion evaluated at the synthesis pump frequency $D_\mathrm{int}(f_\mathrm{sp})$, in the absence of strong overlap between the primary and the synthesized elements, which is our case since the synthesis pump is placed outside of the primary soliton spectral envelope. In Fig.~\ref{fig:OverlapMeasure}d, we compare $\delta f$ and the theoretical value of $D_\mathrm{int}(f_\mathrm{sp})$, and find that they are in good agreement and within the expected Kerr shift (couple of GHz) induced by the synthesis pump.

\noindent\textbf{Probing the relative noise between the two portions of the comb} — Thus far we have shown that the spectrum produced by our dual-pump system is consistent with the picture in which FWM-BS mediates soliton spectral translation, with new DWs generated on both the low and high frequency sides of the spectrum. Moreover, we have explicitly shown that the comb tooth spacing is translated, and there is an overall shift between the two portions of the spectrum. In addition, the narrow heterodyne beat notes across the spectrum suggest that the primary and synthesis portions of the comb each exhibit low noise, but so far, their relative noise has not been experimentally considered. Now, if FWM-BS is indeed the dominant process in the spectral broadening of the primary DKS state into an ultra-broadband comb, we should expect that no added phase noise (other than that of the synthesis laser) will be accrued on the synthesis component teeth, and that there will be low noise between the synthesis portion and the primary portion of the comb. To investigate this, we seek to mix two comb teeth from the overlap region, where one comb tooth comes from the primary portion of the comb and the other tooth comes from the synthesis portion. Because of the large frequency offset $\delta f\approx$ 200~GHz, as observed earlier, we are unable to directly measure such a large beat frequency. Instead, we use an electro-optic modulation (EO) comb, generated through phase modulating an independent 1300~nm laser, to span across $\delta f$ with a measurable (6.0156 GHz) repetition rate (\cref{fig:BeatNoteStitching}a-c), and the beat note of both microcomb teeth with respect to the EO comb is recorded (\cref{fig:BeatNoteStitching}d). As the 1300~nm laser that seeds the EO comb is free-running and not locked to the microcomb (which itself is free running), the obtained beat note is relatively broad. However, by mixing together the two beat notes in a specific configuration, we can suppress this helper laser phase noise. In particular, if the EO comb teeth are correctly positioned on the same side of the DKS comb teeth, for instance both at larger frequency, the two beat notes can be mixed together in order to suppress the phase noise of the 1300~nm helper laser (\cref{fig:BeatNoteStitching}a). The obtained signal after mixing will only correspond to the beating between the two component teeth of the DKS down-shifted by a fixed frequency that is related to the microwave signal generator that drives the EO comb (and which has very low phase noise in comparison to the free-running laser systems). The expected behavior based on an underlying nonlinear process dominated by the FWM-BS is corroborated in experiment, where a narrow single tone, indicating a low-noise state between the two portions of the comb, is observed (\cref{fig:BeatNoteStitching}e). Given that a DKS soliton (the primary portion of the comb) is a low-noise state, this suggests that the spectral translation process has resulted in a synthesis portion that is also low-noise. Though stabilization of the DKS and the pump lasers has not been performed, limiting the ability to draw a final conclusion about the phase coherence between the two portions of the comb, considering the earlier measurements and theoretical analysis, which show that the repetition rate in both portions of the comb is unvarying and identical to within our measurement capability, suggests - as expected from the physical picture of FWM-BS - that the soliton spectral translation process results in a single, fully phase-coherent comb. The frequency of each comb tooth is then known once there is knowledge of the repetition rate, the carrier-envelope offset frequency, and the overlap-offset frequency $\delta f$, whose value is determined by the integrated dispersion value at the synthesis pump.

{\indent}Hence, based on the experimental demonstrations throughout this work, it seems likely that the ultra-broadband microcombs we have demonstrated can be used for metrological purposes, such as $f-2f$ self-referencing, provided that, along with the comb repetition frequency and carrier-envelope offset frequency, the overlap-offset frequency $\delta f$ is measured. In such a scenario, where both pumps can be independently stabilized, the large power available from the synthesis pump would be of particular appeal, as it could be efficiently frequency doubled and, through proper dispersion engineering of the resonator, could be made resonant with the high frequency DW. \\
%
%
%
\noindent\textbf{Discussion} — Zhang and colleagues have recently realized a similar dual pumping configuration, in which an auxiliary laser at 1330~nm spectrally broadens a 1550~nm soliton microcomb down to wavelengths of around 1275~nm~\cite{zhang_spectral_2020}. The spectrum extends from 1275~nm to 1720~nm (taken at the -50~dB points relative to the maximum), an extent of 60~THz, and the auxiliary laser is responsible for about 22~THz of spectral extension on the high frequency side, resulting in a about a factor 1.5$\times$ increase of the bandwidth of the comb. In contrast, here our 1557~nm auxiliary laser causes a broadening of the comb spectrum on both low and high frequency sides of the original soliton spectrum centered at 1063~nm, with a comb extending from 737~nm to 2190~nm, an extent of 270~THz, and the auxiliary laser is responsible for about 93~THz and 79~THz of spectral extension on the low and high frequency sides, respectively (173~THz broadening in total), in our case extending the comb by more than a factor 2.6. Perhaps more important than the characteristics of the comb broadening is its fundamental physical origin. In \cite{zhang_spectral_2020}, spectral broadening is generally attributed to XPM effects, which leaves open many questions about the relationship between the original comb and the spectrally broadened region - in particular, whether any offset between the two regions is indicative of independent frequency combs. Here, we show that the dominant process behind spectral broadening in our system is FWM-BS, so that soliton comb teeth generated by the primary pump are spectrally translated to both the low frequency and high frequency sides of the spectrum. In addition, we show that the overlap offset frequency between the primary region of the comb and the spectrally translated one is inherent to the microring resonator geometry, and is not a signature of two independent frequency combs. In particular, the natural discrepancy between the DKS comb teeth and cavity resonance frequency causes the synthesis pump to be offset from the nearest DKS comb tooth by the value of the integrated dispersion at this pumped mode, which we confirm through measurement of the overlap offset frequency as a function of ring geometry. Finally, we have performed several measurements that indicate that the FWM-BS process in our system directly translates primary DKS comb teeth to another spectral window, resulting in a comb state with low-noise in each portion as well as in the overlap region, pointing to the potential use of such ultra-broadband frequency combs for metrological purposes. This allows us to introduce a simplifying tool for designing this new kind of frequency comb, by summarizing the non-linear interaction and the position of all the DWs generated by the dual-pump system through the synthetic dispersion. Such coherent ultra-broadband frequency combs through DKS spectral translation could find many applications, in particular, by harnessing the high power of the synthesis pump, which would aid in $f$-2$f$ self referencing, for monolithic integration of $f$-3$f$ within a $\chi^{(3)}$ platform, and by pushing the limit of coherent DW generation further into the visible.

\bibliography{Moille_SyntheticDisp}

%
\vspace{3ex}
\noindent \textbf{\large Methods}\\
\textbf{Device design}— We use 775~nm thick Si\textsubscript{3}N\textsubscript{4} ring resonators, which were fabricated at Ligentec SA, with a fixed ring radius of 23~{\textmu}m, a ring width ($RW$) that is varied between 1088~nm and 1140~nm across the devices, and a surrounding silica cladding. The access waveguides for coupling to/from the rings are tapered down to 200~nm at the facets, resulting in about 6~dB and 5~dB insertion losses per facet at 192~THz and 282~THz respectively. We use a pulley waveguide with a width of $W=550$~nm, a length of $L_\mathrm{c}=9$~{\textmu}m, and a gap $G=370$~nm. The expected frequency-dependent coupling, computed using the coupled mode theory formalism developed in ref.~\cite{moille_broadband_2019}, exhibits a resonance-free spectrum and $Q_c$ that varies within one order of magnitude over an octave (Extended Data \cref{fig:coupling}). \\

\noindent \textbf{\large Data availability} \\
The data that supports the plots within this paper and other findings of this study are available from the corresponding authors upon reasonable request.\\

\noindent \textbf{\large Acknowledgments} \\
The ring resonators were fabricated at Ligentec Inc. The authors thanks Alfredo de Rossi and Sylvain Combrié for fruitful discussions. \\

\noindent \textbf{\large Funding} \\
The authors acknowledge funding from the DARPA APHI, DARPA ACES, DARPA DODOS, and NIST-on-a-chip programs. A.R. and X.L. acknowledge support under the Cooperative Research Agreement between the University of Maryland and NIST-PML, Award no. 70NANB10H193.\\

\noindent \textbf{\large Author contributions}\\
G.M developed the theoretical framework, performed the simulations, designed the ring resonators and conducted the experiments. E.P., J.R.S., and T.S.R. helped with the experiments, A.R. helped with ring resonator design, and Y.C. and X.L contributed in the understanding of the physical phenomenon. G.M. and K.S. wrote the manuscript, with input from all authors, and K.S supervised the project. All the authors contributed and discussed the content of this manuscript.\\

\noindent \textbf{\large Competing interests}\\
The authors declare no competing interests.\\

\noindent \textbf{\large Additional Information}\\
Correspondence and requests for materials should be addressed to G.M. and K.S.


\clearpage


\clearpage
\section*{Extended Data}
\beginsupplement

\subsection{Four-Wave Mixing Bragg Scattering Matching Conditions}
\label{sub:sup-mat-FWMBS}
Four-wave mixing (FWM) relies on the basic conditions of energy and momentum conservation, which in a microring resonator translate to frequency ($\omega$) and azimuthal mode number ($M$) conservation. In the case of FWM Bragg Scattering (FWM-BS), the signal photon (which in our case is a comb tooth from the primary portion of the comb, labeled $pDKS$) can be scattered to two spectrally-translated idlers, labeled as $sDKS^{\pm}$, such that these fundamental energy/momentum conservation equations become:

\begin{equation}
\label{sup_eq:FWM}
\begin{cases}
 & \omega_\mathrm{pDKS} \pm \omega_\mathrm{pp} =  \omega_\mathrm{sDKS}^{\pm} \pm \omega_\mathrm{sp}\\
 & M_\mathrm{pDKS} \pm M_\mathrm{pp} = M_\mathrm{sDKS}^{\pm} \pm M_\textsc{sp}
\end{cases}
\end{equation}

The frequency shift in this spectral translation process is set by the difference in the frequencies of the primary pump $pp$ and synthesis pump $sp$, and both a higher frequency idler ($\omega_\mathrm{sDKS}^{+}$) and a lower frequency idler ($\omega_\mathrm{sDKS}^{-}$) relative to the signal can be generated, if they are phase- and frequency-matched~\cite{li_efficient_2016}.

In a multi-mode resonator, such as the ring resonators studied here, the integrated dispersion ($D_\mathrm{int}$) effectively characterizes the discrepancy between the modal frequencies and that of a zero-dispersion resonator, \textit{i.e.}, one in which all resonances are separated by the same free spectral range (FSR). $D_\mathrm{int}$ is usually referenced at the pumped mode, so that we introduce the relative mode number with respect to the pumped mode $\mu = M - M_\mathrm{pp}$ (\textit{i.e} $\mu_\mathrm{pp} = 0$), with:

\begin{equation}
 \label{sup_eq:Dint}
 D_\mathrm{int}(\mu) = \omega_\mathrm{res}(\mu) -  \left(\omega_\mathrm{pp} + D_1 \mu \right)
\end{equation}

\noindent where $\omega_\mathrm{res}(\mu)$ are the cavity resonance frequencies, and $D_1=2\pi FSR$, computed at the primary pumped mode.

From the the above equations, we can compute the frequency of each resonance of interest from the integrated dispersion, assuming that the signal $\mu_\mathrm{pDKS}$ is part of the primary DKS and that the idler $\mu_\mathrm{sDKS}^{\pm}$ is part of the spectrally-translated (synthesized) portion of the DKS, and hence on a fixed frequency grid relative to the synthesis pump.  We then have:
\begin{align}
  & \omega_\mathrm{pDKS} = \omega_\mathrm{pp} + \mu_\mathrm{pDKS} D_1                 \nonumber\\
  & \omega_\mathrm{sp} = \omega_\mathrm{pp} + \mu_\mathrm{sp} D_1  + D_\mathrm{int}(\mu_\mathrm{sp} )            \\
  & \omega _\mathrm{sDKS}^{\pm} = \omega_\mathrm{pp} + \mu_\mathrm{sDKS}^{\pm} D_1 + D_\mathrm{int}(\mu _\mathrm{sDKS}^{\pm}) \nonumber
\end{align}

This allows us to rewrite the energy/momentum conservation condition for FWM-BS of \cref{sup_eq:FWM} in a single equation such that:

\begin{align}
    \label{sup-eq:dint-fwm-general}
    \left(\mu_\mathrm{pDKS} \pm \mu_\mathrm{pp} \right )D_1 =
    & \Big( D_\mathrm{int}(\mu_\mathrm{sDKS}^{\pm}) \pm D_\mathrm{int}(\mu_\mathrm{sp}) \Big) + \nonumber \\ & \left(\mu_\mathrm{sp} \pm \mu_\mathrm{sDKS}^{\pm} \right )D_1
\end{align}

Because of the asymmetric nature of the integrated dispersion (due to odd order dispersion coefficients), both high and low frequency idlers will be obtained for different input signals to respect the above equation. First, we study the case of the lower frequency idlers, as described in Fig 1(a-b). Equation \ref{sup-eq:dint-fwm-general} in this case becomes:
\begin{align}
    \label{sup-eq:dint-fwm-minus}
    \left(\mu_\mathrm{pDKS} - \mu_\mathrm{pp} \right )D_1 = & \Big(D_\mathrm{int}(\mu_\mathrm{sDKS}^{-}) - D_\mathrm{int}(\mu_\mathrm{sp}) \Big) + \nonumber \\ &\left( \mu_\mathrm{sDKS}^{-} - \mu_\mathrm{sp} \right )D_1
\end{align}

For efficient idler generation, we start by assuming that momentum is conserved to allow FWM-BS. In the frequency comb, this is trivial as it involves the simple condition that the number of resonator modes separating the primary pump and the signal $\mu_\mathrm{pDKS} - \mu_\mathrm{pp}$ must be the same as the number of modes separating the synthesis pump and the idler $\mu_\mathrm{sDKS}^{-} - \mu_\mathrm{sp}$. This will be respected due to the large number of comb teeth that can act as the signal. Assuming these two conditions, frequency- and phase-matching for FWM-BS becomes:
\begin{align}
    \label{sup-eq:dint-minus-condition}
    \begin{cases}
        D_\mathrm{int}(\mu_\mathrm{sDKS}^{-}) =  D_\mathrm{int}(\mu_\mathrm{sp}) \\
        \mu_\mathrm{sDKS}^{-} - \mu_\mathrm{sp} = \mu_\mathrm{pDKS} - \mu_\mathrm{pp}
    \end{cases}
\end{align}
This implies that any resonator mode, as long as it is at a lower frequency than the the largest frequency comb tooth, will undergo FWM-BS if it exhibits the same integrated dispersion value as that at the synthesis pump frequency. The power of this idler will be limited by the primary comb tooth that realizes momentum conservation. In our case, we have the synthesis pump at a lower frequency than the primary pump, so that this condition describes both the frequency comb shape around the synthesis pump and the newly created DW at lower frequency.

On the other hand, the higher frequency idler respects the equation:
\begin{align}
    \label{sup-eq:dint-fwm-plus}
     D_\mathrm{int}(\mu_\mathrm{sDKS}^{+}) + \left(\mu_\mathrm{sDKS}^{+} - \mu_\mathrm{pp} \right)D_1 = \nonumber \\ - D_\mathrm{int}(\mu_\mathrm{sp}) + \left(\mu_\mathrm{pDKS} - \mu_\mathrm{sp}  \right)D_1
\end{align}

Momentum conservation will be respected for any spacing between the primary comb tooth and synthesis pump $\mu_\mathrm{pDKS} - \mu_\mathrm{sp}$ that matches the idler and primary pump spacing $\mu_\mathrm{sDKS}^{+} - \mu_\mathrm{pp} $, leading to the condition:

\begin{align}
    \label{sup-eq:dint-plus-condition}
    \begin{cases}
        D_\mathrm{int}(\mu_\mathrm{sDKS}^{+}) =  -D_\mathrm{int}(\mu_\mathrm{sp}) \\
        \mu_\mathrm{sDKS}^{+} - \mu_\mathrm{pp} = \mu_\mathrm{pDKS} - \mu_\mathrm{sp}
    \end{cases}
\end{align}

Therefore, any resonator mode, as long as it is at higher frequency than the the lowest frequency comb tooth, and is equal and opposite in sign to the integrated dispersion value at the synthesis pump frequency will undergo a FWM-BS process. Interestingly, a primary portion DW must occur between the primary pump and the idler, as an integrated dispersion zero-crossing must happen for the above condition to hold.  In Fig.~\ref{fig:ConceptSynthesis}(d), this DW occurs at 358~THz, for example.

\begin{figure}[!t]
 \begin{center}
  \includegraphics[width=\linewidth]{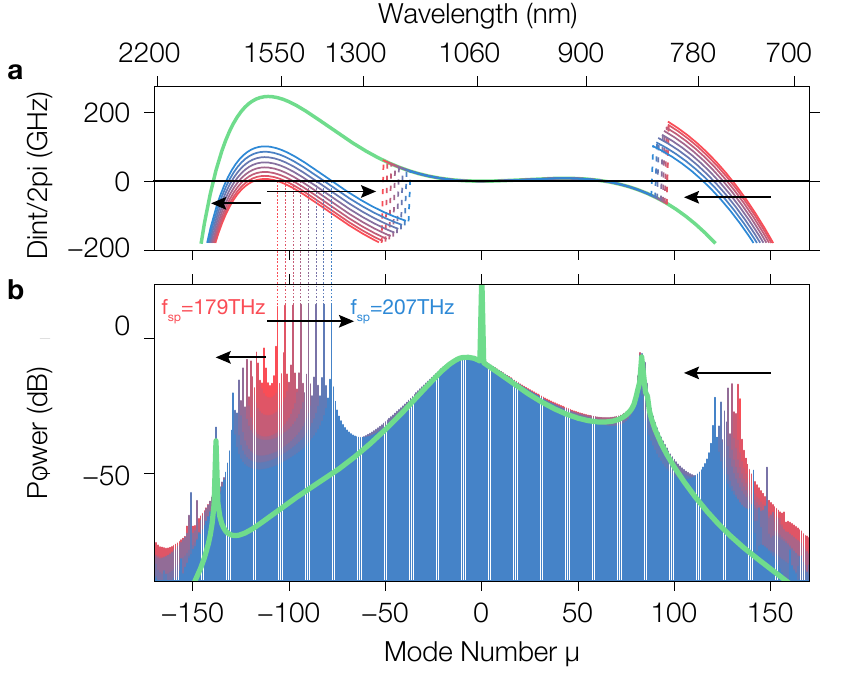}
 \end{center}
 \caption{\label{fig:Supmat_DispersionControl}\textbf{$|$~Tuning of the newly generated DWs by tuning of the synthetic dispersion.} \textbf{a,} Fixed geometric integrated disperson (green) and resulting synthetic dispersion (red to blue) for different synthesis pump frequencies and a fixed primary pump frequency. \textbf{b,} Results of generalized LLE simulations showing how the comb spectrum changes as a function of synthesis pump frequency, and how the position of the newly generated DWs is predicted by the zero crossings of the synthetic dispersion curves in \textbf{a}. The solid green curve shows the spectral envelope when only the primary pump is applied.}
\end{figure}

\subsection{Lugiato Lefever Model}
\label{sub:sup-lle}
The Lugiato-Lefever Equation (LLE) can be derived from basic couple mode theory~\cite{hansson_numerical_2014, chembo_spatiotemporal_2013}, which is similar to the temporal approach using the nonlinear Schr\"odinger equation (NLSE) with the assumption of periodic boundary conditions and a slowly varying envelope~\cite{coen_modeling_2012}. In many treatments, the detuning of the pump is simplified and introduced as a linear term in the equation, making it convenient for solving a single pump system. However, it has been demonstrated in ref.~\cite{taheri_optical_2017} that avoiding such approximation allows for solving a system under $N$ driving forces. Instead of choosing the frequency grid of the resonant modes with an origin defined by a single pump, the grid is determined by the spectral domain under study (with the origin placed at the center of the grid). The generalized LLE becomes:
\begin{align}
 \label{sup-eq:LLE}
 \frac{\partial A(t, \tau)}{\partial t} = & \left(-\frac{\alpha}{2} - i\gamma L |A|^2 \right)A                                          \\
 +                                        & i t_\mathrm{r} \mathrm{FT}^{-1}\left[D_\mathrm{int}^{0}\tilde{A}(t,\omega)\right] \nonumber \\
 +                                        &
 \sum_\mathrm{p}\kappa_\mathrm{ext}|_\mathrm{p} \mathcal{F}_p  \mathrm{e}^{i\sigma_\mathrm{p}} \nonumber
\end{align}

\noindent where the subscript $p$ refers to either the synthesis pump or the primary pump ($p=\{\mathrm{pp}; \mathrm{sp}\}$), the index $0$ represents the center of the simulation domain and hence $D_\mathrm{int}^{0}$ is the integrated dispersion of the resonator computed at the center of the simulation domain and $\mu_0 = 0$, $\tilde{A}(t, \omega) =\mathrm{FT}\left[{A}(t, \tau)\right]$ is the Fourier Transform of the field,  $\mathcal{F}_p = -i\sqrt{P_\mathrm{in_p}}$ represents the $p$th driving field (either the primary or synthesis pump), $\sigma_\mathrm{p} = \delta\omega_p t + D_\mathrm{int}^{0}(\mu_p)t - \mu_p\theta$, represents the phase shift of the $p$ driving field, from which it is obvious that the part relative to the fast time (or resonator angle $\theta$) corresponds to the spectral shift away from the center of the simulation domain, and the slow time $t$ phase corresponds to the dispersion shift away from the uniform spectral grid and the relative detuning of the $p$ pump from its pumped mode, $\kappa_\mathrm{ext}|_\mathrm{p}$ is the waveguide-resonator coupling rate for  the $p$ pumped mode,  $\alpha$ is the total loss of the resonator, $\gamma$ is the non-linear coefficient (here assumed constant), $L$ is the resonator round-trip length, and $t_\mathrm{r}$ is the round trip time. We note that the synthetic dispersion model replaces $D_\mathrm{int}^{0}$ with $D_\mathrm{synth}$ and a single pump field for a symmetric domain.

\begin{figure}[!t]
 \begin{center}
  \includegraphics[width=\linewidth]{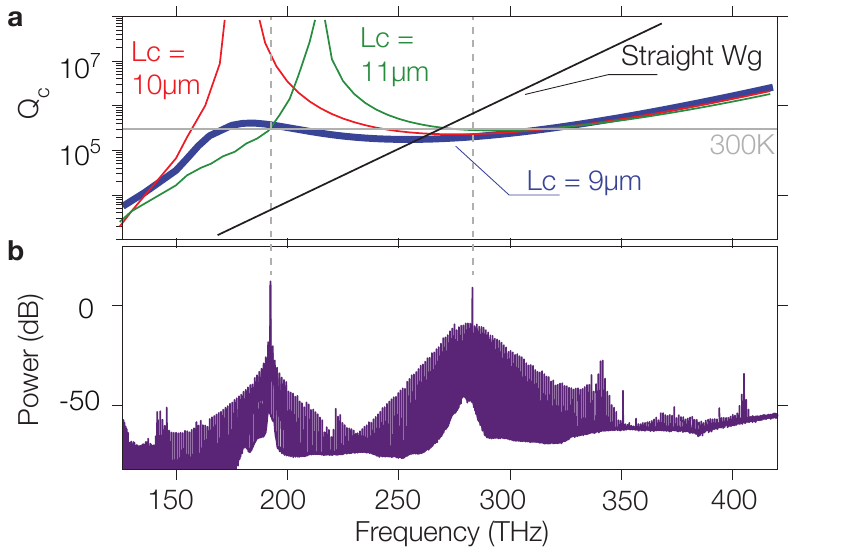}
 \end{center}
 \caption{\label{fig:coupling} \textbf{$|$~Pulley scheme for broadband resonator-waveguide coupling}. \textbf{a,} Calculated coupling quality factor $Q_\mathrm{c}$ for a waveguide width = 550~nm and a ring resonator with $RR=23$~{\textmu}m and $RW=1100$~nm, for three different pulley lengths $L_\mathrm{c}=\{9$, $10$, $11$\}~{\textmu}m. $L_\mathrm{c}=9$~$\mu$m provides nearly equal $Q_\mathrm{c}$ at the primary and synthesis pump frequencies and limited variation in $Q_\mathrm{c}$ across the full comb spectrum, In contrast, conventional straight coupling, shown in black, results in severe overcoupling at the synthesis pump wavelength and severe undercoupling at high frequencies. \textbf{b,} Experimental ultra-broadband DKS spectrum for $RW=1103$~nm, illustrating the different regions of interest for in-coupling and out-coupling to/from the ring resonator. 0~dB is referenced to 1~mW, i.e., dBm.}
\end{figure}

\subsection{Dispersive Wave Tuning Independent of Geometric and Material Dispersion}

In this section we further justify use of the term 'synthetic dispersion' to capture the net effects of the new nonlinear-wave mixing processes that occur in the dual pump system. Usually, dispersion is driven by two main components. The first is material dispersion, where the chromatic dependence is such that the dispersion becomes more normal the closer the wavelength is to the band-gap, and is a significant challenge to reaching short wavelengths in integrated frequency combs. The second is geometric dispersion, where the wavelength-dependent confinement of the light within a guided mode geometry provides a modification of the phase velocity of light different from the bulk material trend, therefore providing a counter-balance to material dispersion. Thus, for a fixed platform (\textit{i.e.} a chosen set of photonic materials), the typical view is that the geometry must be changed in order to modify the dispersion of the resonator, and hence the resulting spectral shape of the generated DKS. In this work, we have shown that by introducing the synthesis pump, we provide a new degree of freedom in controlling the comb spectrum. Apart from the significant spectral broadening mediated by the FWM-BS processes we have described, control of the synthesis pump frequency results in tuning of the DWs (~\cref{fig:Supmat_DispersionControl}b). This is a consequence of a change in the synthesis pumped mode resulting in new modes that match the FWM process. The net result is that DW positions are modified without requiring modification of the geometry or material. This tuning of the DW positions with synthesis pump frequency is well-reproduced by the synthetic dispersion (~\cref{fig:Supmat_DispersionControl}a), whose zero-crossings closely matched the DW positions.

\subsection{Coupling Design}
\label{sub:sup-coupling}
In order to in-couple two pumps separated by close to 100~THz and out-couple close to a 300~THz bandwidth comb, straight waveguide coupling cannot be used as it exhibits a coupling quality factor $Q_\mathrm{c}$ (inversely proportional to the coupling rate $\kappa_\mathrm{ext}$) varying by orders of magnitude over an octave~\cite{moille_broadband_2019}. However, pulley couplers exhibiting phase mismatch can achieve much more spectrally flat coupling over the bandwidth of interest. In our system, we use a pulley waveguide with a width of $W=550$~nm, a length of $L_\mathrm{c}=9$~{\textmu}m, and a gap $G=370$~nm. Such coupling, computed using the coupled mode theory (CMT) formalism developed in ref.~\cite{moille_broadband_2019}, exhibits a resonance-free spectrum and $Q_c$ that varies within one order of magnitude over an octave (\cref{fig:coupling}). This allows for efficient extraction of the whole ultra-broadband dual-pumped frequency comb without any gap in the spectrum, thanks to the resonance-less coupling. However, this design is quite sensitive to phase-mismatch, and as a result the geometric parameters that define the coupling, as longer pulley lengths would exhibit coupling resonances that either limit coupling of the long wavelength pump (case $L_c=10$~{\textmu}m) or the spectral region in between the primary pump and the synthesis pump would (case $L_c=11$~{\textmu}m). However, the optimal flat coupling over a large bandwidth comes at the cost that the coupling quality factor $Q_c$ is lower than the intrinsic quality factor ($Q_i \approx 10^6$), resulting in over-coupled devices.

\begin{figure}[!t]
 \begin{center}
  \includegraphics[width=\linewidth]{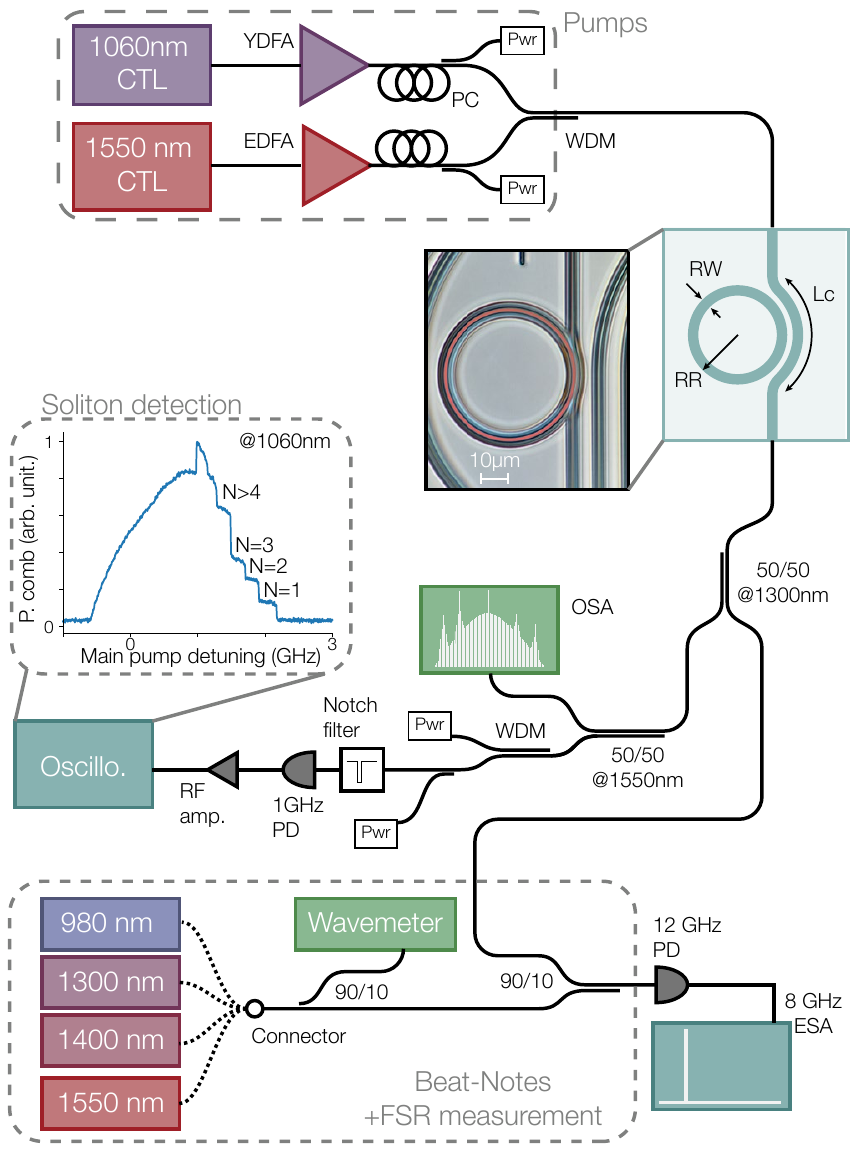}
 \end{center}
 \caption{\label{fig:setup}\textbf{$|$~Experimental setup.} Two amplified lasers at 282~THz (1063~nm) and 192~THz (1557~nm) pump a microring resonator. The output signal is split between different paths that allow the observation of the spectrum with an OSA and the comb power exhibiting the DKS step signature (experimental data labeled as `Soliton detection’). One output path is used to measure the DKS comb tooth frequency through beat-note measurement with different frequency lasers and a wavemeter for absolute wavelength measurement. Pwr: Powermeter, CTL: Continuous Tunable Laser, PC: Polarization Controller, PD: PhotoDiode, OSA: Optical Spectrum Analyzer, ESA: Electrical Spectrum Analyzer, WDM: Wavelength Demultiplexer.}
\end{figure}

\begin{figure*}[!t]
 \begin{center}
  \includegraphics[width=\linewidth]{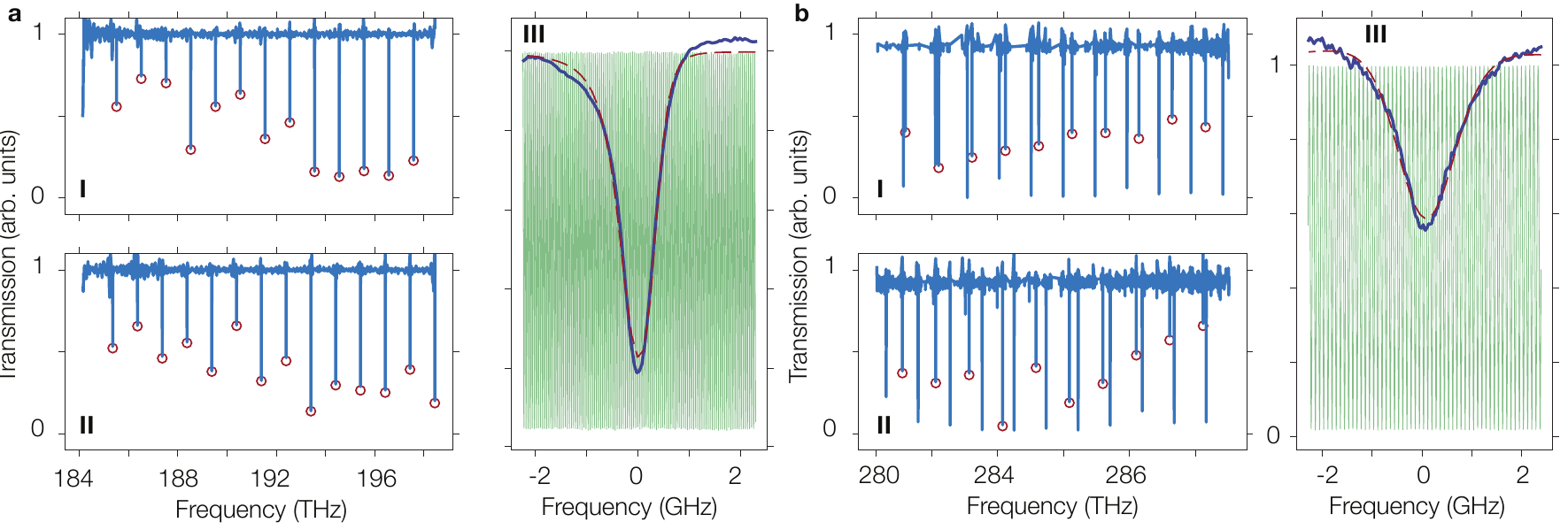}
 \end{center}
 \caption{\label{fig:linear}\textbf{$|$~Linear measurement of the ring resonator}. \textbf{a,} Transmission spectrum around 1550~nm for $RW=1088$ (I) and $1103$~nm (II), with a zoom-in of a representative resonance highlighted by the black dashed line in II displayed in panel III, with the nonlinear least squares fit shown in the red dashed lined. The 42~MHz MZI that serves as a calibration is displayed in green. \textbf{b,} Transmission spectrum around 1050~nm for the same resonators, namely $RW=1088$ (I) and $1103$~nm (II), with the fit of a representative resonance highlighted by the black dashed line in II. The nonlinear least squares fit to the zoomed-in data is shown as a red dashed lined in III. The 83~MHz MZI that serves as a calibration for the 1060nm band is displayed in green.}
\end{figure*}

\subsection{Characterization setup}

The characterization setup used in measurement of the microcomb devices is shown in \cref{fig:setup}. The resonator is pumped in the transverse electric (TE) polarization in two bands, with a primary pump at 1063~nm amplified by a ytterbium-doped fiber amplifier and a synthesis pump at 1557~nm amplified with an erbium-doped fiber amplifier, which are both coupled to the chip using a wavelength division multiplexer and a lensed fiber. The on-chip primary pump power, \textit{i.e.} at 282~THz, is $P_\mathrm{pp}=200$~mW (7.5~dB insertion loss per facet) and the on-chip synthesis pump power $P_\mathrm{sp}=250$~mW (6~dB insertion loss per facet). The generated microcomb is out-coupled and split between an optical spectrum analyzer and another path allowing for beat-note measurements and comb-power detection. As expected with the synthesis pump scheme, appropriate choice of the frequency of each laser thermally stabilizes the ring resonator and results in a relatively pure Kerr response (i.e., thermal effects mitigated) with a clear signature of soliton steps for different soliton orders (inset to~\cref{fig:setup}), ultimately reaching the lowest order single soliton state.\\

\subsection{Linear Measurements}

Linear measurements to determine the mode family and the quality factor of the ring resonator, in both the 1550 and 1060~nm band, are performed by sweeping a continuous tunable laser (CTL) and recording the transmission spectrum (\cref{fig:linear}). A Mach-Zehnder Interferometer (MZI), with a free spectral range of 42~MHz and 83~MHz for the 1550~nm and 1060~nm band respectively, is used to calibrate the laser sweeps. This allows retrieval of the quality factor of the resonances in both bands, which are in the over-coupled regime as expected from the broadband-flat coupling described in the previous section.The modes in the spectra shown in \cref{fig:linear} exhibit $Q_\mathrm{i}\approx1.1\times10^6$ and $Q_\mathrm{c}\approx 3.5\times10^5$ for the average intrinsic and coupled quality factors, respectively.

\end{document}